\documentclass{emulateapj}
\usepackage{amsmath}
\usepackage{multirow}
\usepackage{url}
\shorttitle{High energy emission from TDEs}
\begin{document}

\title{Search for high energy gamma-ray emission from tidal disruption events with  the \textsl{Fermi} Large Area Telescope
 }

\author{Fang-Kun Peng\altaffilmark{1,2}, Qing-Wen Tang\altaffilmark{3} and Xiang-Yu Wang\altaffilmark{1,2}
}
\altaffiltext{1}{School of Astronomy and Space Science, Nanjing University, Nanjing 210093, China; xywang@nju.edu.cn}
\altaffiltext{2}{Key laboratory of Modern Astronomy and Astrophysics (Nanjing University), Ministry of Education, Nanjing 210093, China}
\altaffiltext{3}{School of Science, Nanchang University,Nanchang 330031, China}

\begin{abstract}
Massive black holes at galaxy center may tear apart a star when
the star passes occasionally within the disruption  radius, which
is the so-called tidal disruption event(TDE). Most TDEs radiate
with thermal emission resulted from the accretion disk, but three
TDEs have been detected in bright non-thermal X-ray emission,
which is interpreted as arising from the relativistic jets. Search
for high-energy gamma-ray emission from one relativistic TDE
(Swift J164449.3+573451) with the \textsl{Fermi} Large Area
Telescope (LAT) has yielded non-detection. In this paper, we
report the search for high energy emission from the other two
relativistic TDEs (Swift J2058.4+0516 Swift J1112.2-8238) during
the flare period. No significant GeV emission is found, with an
upper limit fluence in LAT energy range being less than $1\%$ of
that in X-rays. Compared with gamma-ray bursts (GRBs) and blazars,
these TDEs have the lowest flux ratio between GeV emission and
X-ray emission. {The non-detection of high-energy emission from
relativistic TDEs could be due to that the high-energy emission is
absorbed by soft photons in the source. Based on this hypothesis,
upper limits on the bulk Lorentz factors,  $\Gamma\la 30$, are
then obtained for the jets in these TDEs.} We also search for
high-energy gamma-ray emission from the nearest TDE discovered to
date, ASASSN-14li. No significant GeV emission is found and an
upper limit of $L(\rm 0.1-10 GeV)\le 4.4\times 10^{42}$ erg
s$^{-1}$ (at $95\%$ confidence level) is obtained for the first
$10^{7}$ s after the disruption.

\end{abstract}


\section{Introduction}
TDE is an astronomical phenomenon that occurs when a star gets too
close to  a supermassive black hole in the galaxy center and is
disrupted by the tidal force of the black hole. Part of stellar
material is bound and accreted by the central black hole,
resulting in a bright optical, UV and soft X-ray emission
\citep[and references
therein]{1988Natur.333..523R,2015JHEAp...7..158L}. There are
growing number of candidate TDEs being discovered in soft X-ray,
ultraviolet and optical surveys, see \citet{2015JHEAp...7..148K}
for a recent review. Recently, three unusual TDE candidates are
discovered by \textsl{Swift},  i.e. Swift J164449.3+573451 Swift
J2058.4+0516 Swift J1112.2-8238 (hereafter  Sw J1644+57,   Sw
J2058+05 and Sw J1112-82 for short respectively), which have very
bright non-thermal hard X-ray and radio emissions
\citep{2011Natur.476..421B,2011Sci...333..203B,2011Natur.476..425Z,2011ATel.3384....1K,2012ApJ...753...77C,
2011ATel.3463....1K,2015MNRAS.452.4297B}.  The luminous
non-thermal X-ray and radio emissions are thought to be produced
by relativistic jets
\citep{2011Sci...333..203B,2011Sci...333..199L,2011Natur.476..421B,2011Natur.476..425Z,2012MNRAS.420.3528M,2012ApJ...761..111C,2014ApJ...788...32W,2015ApJ...798...13L}.
Sw J1644+57 shows a highly variable light curve in X-rays, as
observed by X-Ray Telescope (XRT) onboard \textsl{Swift}. At
redshift $z=0.354$, the isotropic luminosity of the X-ray emission
is as high as $10^{48}-10^{49}$ erg s$^{-1}$. Sw J2058+05 exhibits a
luminous, long-lived X-ray outburst with an isotropic peak
luminosity of $3\times10^{47}{\rm erg\, s^{-1}}$ (at redshift
$z=1.1853$). Its total isotropic energy (0.3-10 keV) on timescale
of the first two months amounts to $10^{54}$ erg. Sw J1112-82 was
initially also discovered by Swift/BAT (Burst Alert Telescope) in
June 2011 as an unknown, long-lived (order of days) $\gamma$-ray
transient source. It exhibits a similar bright X-ray flare and its
position is consistent with the nucleus of a faint galaxy  at
$z=0.89$ \citep{2015MNRAS.452.4297B}. The peak X/$\gamma$-ray
luminosity of Sw J1112-82 exceeds $10^{47}{\rm erg \,s^{-1}}$.

The non-thermal X-ray emission is thought to be produced by
synchrotron radiation of relativistic electrons. One would
naturally expect inverse-Compton scattering emission from the same
electrons, which may produce high-energy gamma-ray emission. As
the three TDEs with relativistic jets emit a total isotropic
energy of about $10^{54}$ erg in X-ray band, comparable to or even
larger than that in the GRB prompt emission and blazars flares,
one would expect the high-energy gamma-ray emission detectable by
\textsl{Fermi}/LAT. Motivated by this, we search for high-energy
gamma-ray emission from these three relativistic TDEs.

ASASSN-14li is a  normal optically-discovered TDE  at distance of
about 90 Mpc \citep{2016MNRAS.455.2918H}.  Transient radio
emission has been detected from this event and modeling of the
radio emission gives a kinetic energy of $10^{48}$ erg in a
non-relativistic or mildly relativistic outflow
\citep{2015arXiv151001226A,2015arXiv151108803V}. The relativistic
electrons producing radio emission may in principle also produce
high-energy gamma-ray emission, although the flux level depends on
the energy of these electrons and the strength of the magnetic
field. It is also proposed that the unbound debris  after the
disruption will encounter the interstellar medium or dense
molecular clouds around the central massive black hole, producing
high-energy $\gamma $-ray afterglow through $pp$ collisions
\citep{2006ApJ...645.1138C,2007A&A...473..351C,2015arXiv151206124C}.
Thus, we also search for high-energy gamma-ray emission from the
normal TDE ASASSN-14li.

The analysis of the \textsl{Fermi} LAT  data of TDEs is presented
in Section 2. We present the result for each TDE in Section 3.
Furthermore, we compare the high-energy gamma-ray emission of
relativistic TDEs with that of GRB prompt emission and blazar
flares. Then in Section 4, we derive constraints on the bulk
motion Lorentz factor of the source based on the non-detection of
the high energy emission. Finally we give a short summary.
Throughout this paper, we take the standard $\Lambda$CDM cosmology
with parameters $H_0 =71$ km s$^{-1}$ Mpc$^{-1}$, $\Omega_m=0.27$,
and $\Omega_{\Lambda}=0.73$.

\section{Data}
\subsection{$\gamma$-ray data }
The \textsl{Fermi} LAT is a pair conversion telescope designed to
cover the energy band from 20 MeV to greater than 300 GeV, and
operates primarily in an all-sky scanning survey mode
\citep{2009ApJ...697.1071A,2009APh....32..193A}. The LAT data
currently being released by the FSSC (\textsl{Fermi} Science
Support
Center\footnote{\url{http://fermi.gsfc.nasa.gov/ssc/data/}}) have
been processed using the "Pass 8" (P8R2) event-level analysis, and
specifically P8R2 source event class data are selected here. The
analysis is based on the LAT science tools version v10r0p5. In
order to reduce the contamination from Earth Limb emission, less
than $52^{\circ}$ of LAT rocking angle and less than $90^{\circ}$
of local zenith angle are required. Moreover, we exclude the time
period when the spacecraft is above the South Atlantic Anomaly. We
perform an unbinned maximum likelihood analysis of the selected
data with the following method. Front-back converting photons
(evtype =3) of energies between 100 MeV and 10 GeV are fitted with a
power law spectrum ($N(E)=N_0 (E/E_0)^{-\Gamma}$). All events in a
region of interest (ROI) of $10^{\circ} $ have been used, and
point sources and extending sources within extra $5^{\circ}$ in
the LAT 4-year Point Source Catalog
(3FGL)\citep{2015ApJS..218...23A} are added to the model file. The
galactic diffuse and isotropic emission are molded with
gll\_iem\_v06.fits and iso\_P8R2\_SOURCE\_V6\_v06.txt
respectively. When it leaves too many free parameters to gain a
good spectral fit for a faint source at a short time interval, we
fix the spectral form and free the normalization ('prefactor') of
faint or distant background sources.

For the purpose of comparison, we also analyze the
\textsl{Fermi}/LAT data of two typical blazars, i.e. Mrk 421 and
3C 279, during some flare episodes
\citep{2015arXiv151106851B,2012ApJ...754..114H}. Analysis threads
are similar to the above method. For GRBs, we select the first
\textsl{Fermi} LAT GRB Catalog \citep{2013ApJS..209...11A} as our
sample for comparison with TDEs. The processing procedure for the
LAT data of GRBs follows our previous work
\citep{2015ApJ...806..194T}.

\subsection{X-ray data }
Thanks to \textsl{Swift}/BAT
\citep{2004ApJ...611.1005G,2005SSRv..120..143B}, for  bright X-ray
sources such as the three relativistic TDEs mentioned above, we
can get the daily average count rate in the survey mode (data
quality flag = 0) \citep{2013ApJS..209...14K}
\footnote{\url{http://swift.gsfc.nasa.gov/results/transients/}}
(Because Sw J1644+57 was mistaken for a GRB,  it was observed by
BAT burst mode and had a good follow-up observation
\citep{2011Natur.476..421B}). Considering the limited BAT energy
band (15-50 keV), the corresponding flux will be displayed in the
SED plot (Figure \ref{sed3}) as one data point.

\section{Results}

\subsubsection{Sw J1644+57}
The likelihood analysis of Sw J1644+57 centered at the position
(RA., Dec.)=($251.205^{\circ}, 57.5810^{\circ}$)  results in a low
TS value ($<10$), i.e. no significant high-energy emission is
found from the BAT trigger time to one hundred days after the
trigger ($T_0 +10^7$s). Its $95\%$ confidence upper limit fluxes
are presented in Table \ref{lc3bins} for different time intervals
\footnote{LAT upper limit flux is at 95\% confidence level if not
explicitly specified.}. We fix the spectral photon index  at
$\Gamma=-2.0$. Taking different spectral index $\Gamma$ will
result in a slight, but insignificant difference. We examine the
archival database day by day for three days before $T_0$ and yield
non detection either. Our results are consistent with the results
in \citet{2011Natur.476..421B}, in spite of some slight difference
due to new data (PASS 8), different photon index $\Gamma $ and
background model. The SED in 0.3-150 keV during the bright period
is extracted from \citet{2011Natur.476..421B} and plotted in
Figure \ref{sed3}.  The upper limit  fluence in 0.1-10 GeV is
$F(0.1-10{\rm GeV})\le6.30\times 10^{-6}$ erg cm$^{-2}$ during the
first three days when the source is in the brightest phase in BAT.

Integrating  the BAT emission over the significant emission
period, we get a total 15-50 keV fluence of $1.58\pm
0.1\times10^{-4}$ erg cm$^{-2}$, as shown in Table
\ref{fluencecom}. In order to compare with GRBs, whose fluences
are given in the \textsl{Fermi} Gamma-ray Burst Monitor (GBM) band
(10-1000 keV), we adopt a power law spectrum with an index $-2.15$
($N(E)=N_0 (E/E_0)^{-2.15}$) to extrapolate the 15-50 keV fluence
to the 10-1000 keV range\footnote{As there is no spectral
information of TDEs in the range of 10-1000 keV, we check the
difference when assuming different photon indices. Taking photon
index of -1.80, -2.00, -2.15, -2.30 gives an extrapolation factor
of 5.12, 3.82, 3.21, 2.79 respectively. As the spectrum gets
softening towards the higher energy, as seen in Figure \ref{sed3}
for Sw J1644+57, we adopt -2.15 for the estimation. }, resulting
in a 10-1000 keV fluence of about $5.06\times10^{-4}{\rm erg\,
cm^{-2}}$. One can see that the X/$\gamma$-ray fluence  is about
two orders of magnitude larger than that in 0.1-10 GeV gamma-rays.

\subsubsection{Sw J2058+05}
\begin{figure}
\centering
\includegraphics[angle=0,scale=0.55]{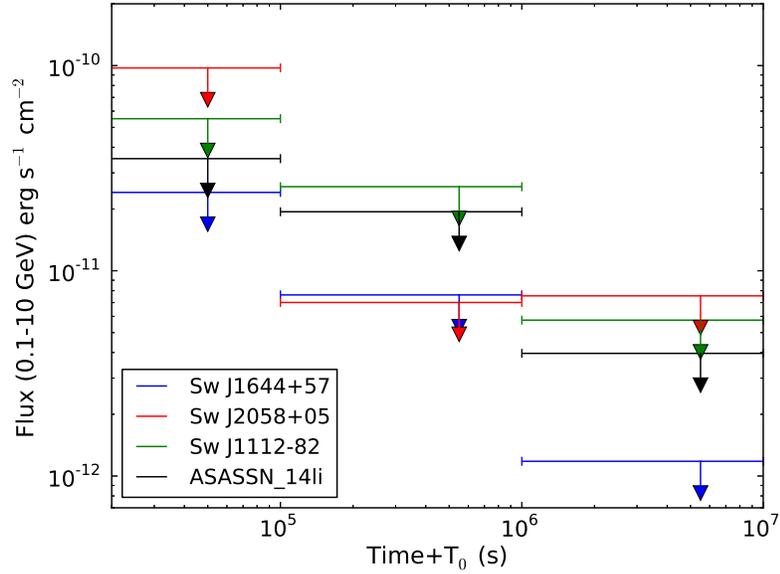}
\caption{\textsl{Fermi}/LAT (0.1-10 GeV) light curves  of
four tidal disruption events. The upper limits are at the 95\%
confidence level.} \label{lightcurve4}
\end{figure}

\begin{figure}
\centering
\includegraphics[angle=0,scale=0.55]{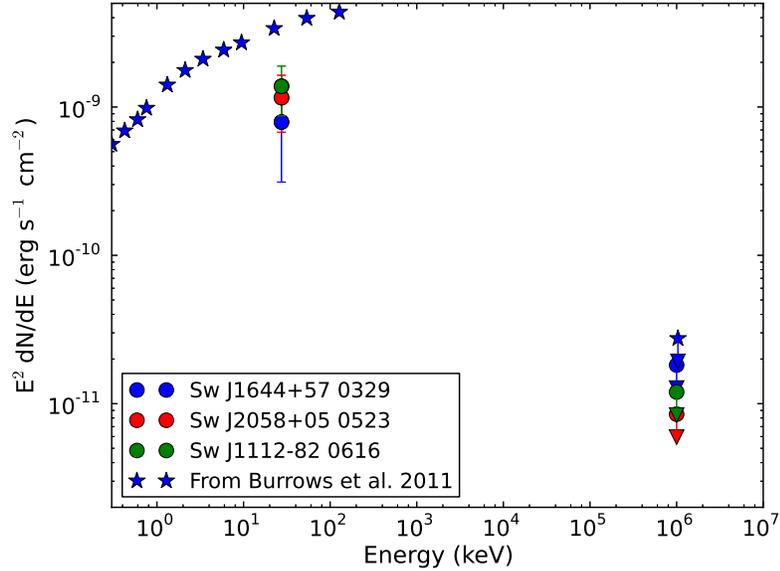}
\caption{The spectral energy distribution of three relativistic
tidal disruption events.  The legend "Sw J1644+57 0329" means the spectrum on 29 March 2011. Data marked with blue stars are taken from \citet{2011Natur.476..421B}.}
\label{sed3}
\end{figure}

The standard gtlike analysis of the \textsl{Fermi}/LAT data of Sw J2058+05
centered at the position (RA., Dec.)=($314.5830^{\circ},
5.2260^{\circ}$) gives a non detection. The upper limit fluxes
from the BAT trigger time to one hundred days after trigger ($T_0
+10^7$s),  divided into three time intervals, are given in Table
\ref{lc3bins} and also shown in Figure \ref{lightcurve4}. Using
the BAT and LAT observational data, the SED on 23 May 2011, which
is the brightest epoch in BAT, is shown in Figure \ref{sed3}.  Sw
J2058+05 has a seventeen-day activity in the BAT observation  with
a total fluence of $\sim 1.64\times 10^{-3}$ erg cm$^{-2}$ when
extrapolated to the energy range of 10-1000 keV (see section
3.0.1). The corresponding LAT upper limit fluence is $F(0.1-10{\rm
GeV})\le 1.67 \times 10^{-5}$ erg cm$^{-2}$ in the same time
interval.

\subsubsection{Sw J1112-82}
The likelihood analysis of the \textsl{Fermi}/LAT data of Sw
J1112-82 centered at (RA., Dec.)=($167.949^{\circ},
-82.6460^{\circ}$) also yields a non-detection.  The upper limit
fluxes from the BAT trigger to one hundred days after trigger
($T_0 +10^7$s) are given in Table \ref{lc3bins} and also shown in
Figure \ref{lightcurve4}. We show the BAT and LAT data on 16 June
2011 (the brightest epoch in BAT) in the SED  plot (Figure
\ref{sed3}). The LAT upper limit fluence during the period with
bright BAT emission (from 16 June 2011 to 19 June 2011) is $9.00
\times 10^{-6}$ erg cm$^{-2}$, while the 10-1000 keV fluence in
the same time interval is $\sim 9.61 \times 10^{-4}$ erg
cm$^{-2}$.

\subsubsection{ASASSN-14li}
The nearest TDE  discovered to date, ASASSN-14li,  is centered at
(RA., Dec.)=($192.063^{\circ},17.7739^{\circ}$)
\citep{2014ATel.6777....1J}. We analyze the \textsl{Fermi}-LAT
data during the period from 22 November 2014 to 22 November 2015
and  find no evidence of high-energy gamma-ray emission, with a
low TS value ($< 10$). The upper limit fluxes are presented in
Table \ref{lc3bins} and also shown in Figure \ref{lightcurve4}.
Search for high energy gamma-ray emission from the same source in
the archival data also yields only upper limit. At distance of 90
Mpc, the upper limit luminosity in 0.1-10 GeV is $L(0.1-10 {\rm
GeV})\le 4.40\times 10^{42}$ erg s$^{-1}$ during the period from
22 November 2014 to 17 March 2015. This limit luminosity is lower
than the soft X-ray (0.1-3 keV) luminosity, which is about
$10^{43}{\rm erg}$ s$^{-1}$ \citep{2016MNRAS.455.2918H}.  However,
it is higher than the predicted gamma-ray luminosity,
$10^{39}-10^{40}$ erg s$^{-1}$, by \citet{2007A&A...473..351C}, so
closer TDEs detected in future are needed to test the prediction.

\subsubsection{Comparison with GRBs and Blazars }
\begin{figure}
\centering
\includegraphics[angle=0,scale=0.55]{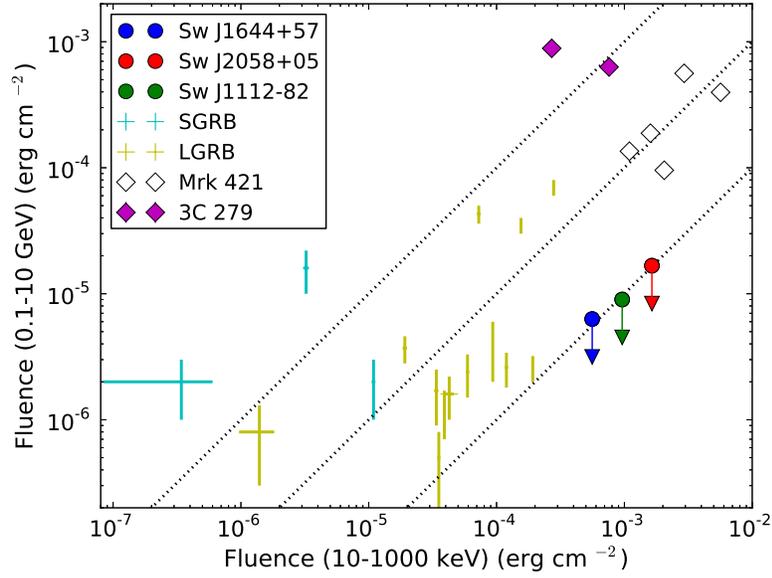}
\caption{Fluence measured by \textsl{Fermi} LAT vs. the fluence
measured by \textsl{Swift}/BAT or  \textsl{Fermi}/GBM during the
X/$\gamma$-ray flare periods for relativistic sources: TDEs, GRBs
and blazars. SGRB and LGRB mean short GRB and long GRB
respectively. The three dished lines from left to right denote the
100\%, 10\%, and 1\% fluence ratio between 0.1-10 GeV and 10-1000
keV.} \label{comparison3}
\end{figure}

Since relativistic TDEs have relativistic jets similar to GRBs and
blazars, we compare the GeV emission in TDEs with the other two
relativistic sources. For GRBs, we make the combined GBM-LAT
spectral analysis during the prompt emission period, and the data
of the GeV fluence and X/$\gamma$-ray fluence are presented in
Figure \ref{comparison3}, which are consistent with the results in
the \textsl{Fermi} LAT First Gamma-Ray Burst Catalog (see their
Fig. 17) \citep{2013ApJS..209...11A}. For blazars, we choose Mrk
421 and 3C 279 as  representatives of BL Lac objects and Flat
Spectrum Radio Quasar (FSRQs) respectively. We select the time
intervals that X-ray emission flares to analyze the \textsl{Fermi}
LAT data \citep{2015arXiv151106851B,2012ApJ...754..114H}. The
modeling of 0.1-10 GeV spectrum of Mrk 421 with a simple power law
function gives  a  photon index of $\Gamma
> -2$, whereas it  gives a soft photon index with $\Gamma <-2$ for 3C 279,
which are consistent with the statistic characteristic of joint
BAT-LAT spectral properties of blazars
\citep{2009arXiv0912.2721S}. Long-time  0.1-300 GeV data gives a
preferred spectral model logarithmic parabola for 3C279 in 3FGL,
so we check our result with this spectral model and find almost no
difference in the fluence. The results of their X/$\gamma$-ray and
GeV fluences are listed in Table \ref{fluencecom} \footnote{The
SED is different for different blazars and different flare states
of a source, thus a simple extrapolation factor ($N(E)=N_0
(E/E_0)^{-2.15}$) may not be appropriate, but anyway, a few times
change of X/$\gamma$-ray (10-1000 keV) fluence will not make the
fluence ratio below 1\% in Figure \ref{comparison3}.}.

In Figure \ref{comparison3}, we compare the X/$\gamma$-ray fluence
and GeV fluence  of the three types of relativistic sources. We
find that all GRBs have  GeV fluences larger than 1\% of the
X/$\gamma$-ray fluences. Among them, a significant fraction of
GRBs have GeV fluences larger than 10\% of the X/$\gamma$-ray
fluence. Blazar flares have similar properties and all of them
have  GeV fluences larger than a few per cent of the
X/$\gamma$-ray fluences. In contrast,  the GeV fluences of all
three relativistic TDEs are smaller than 1\% of the X/$\gamma$-ray
fluences. This {\bf could be due to} that the high-energy
gamma-ray emission in TDEs are highly absorbed, as we will discuss
below.

\section{Discussions}
The non-thermal X-ray emission in three relativistic TDEs may be
produced by relativistic electrons via synchrotron radiation
\citep{2011Natur.476..421B,2012MNRAS.421..908W}. The size of the
source is estimated to be $l'=\Gamma ct_v=3\times10^{13}{\rm
cm}\Gamma_1 t_{v,2}$, where $\Gamma$ is the bulk Lorentz factor of
the source, $t_v$ is the variability timescale of X-ray emission
and we have assumed that the jet aligns with the light of sight of
the observer. Here we use c.g.s. units and the denotation $Q=10^x
Q_x$ in the paper. The observed minimum variability time of the
X-ray emission in Sw J1644+17 is about $t_v=100{\rm s}$
\citep{2011Sci...333..203B,2011Natur.476..421B}. Assuming that the
magnetic field energy density is in equipartition with the
radiation energy density, the magnetic field in the comoving frame
of the source is estimated to be $B'=(8\pi \epsilon_B L_X/4\pi
l'^2\Gamma^4 c)^{1/2}=3\times10^3 {\rm G}\,\epsilon_{B}^{1/2}
L_{X,48}^{1/2}\Gamma_{1}^{-3}t_{v,2}^{-1}$, where $\epsilon_{B}$
is the equipartition factor. For such a magnetic field, the X-ray
photons with frequency $\nu_X=2\times10^{17}{\rm Hz}$ are produced
by relativistic electrons with Lorentz factors of
$\gamma_e=(\frac{2\pi m_e c \nu_X}{\Gamma e
B'})^{1/2}\simeq1.5\times10^3\Gamma_1\epsilon_{B}^{-1/4}L_{X,48}^{-1/4}t_{v,2}^{1/2}$.
For these electrons, we expect an inverse-Compton (IC) component
peaking at $h\nu_{IC}\simeq2\gamma_e^2 h\nu_X\simeq 5 {\rm GeV}$
when other parameters are taken with typical values. The IC flux
should be comparable to or larger than that of the synchrotron
component if $\epsilon_B\la 1$, thus we would expect a GeV
component with a luminosity of $L_{\rm GeV}\ga L_X\sim10^{48}$ erg
s$^{-1}$. The non-detection of such a high-energy component can be
attributed to a high absorption opacity due to low-energy photons
in the source, i.e. the $\gamma\gamma$ absorption optical depth
should be $\tau_{\gamma\gamma}(5 {\rm GeV})\ge 1$. The optical
depth for $\gamma\gamma$ absorption is given by
$\tau_{\gamma\gamma}=\sigma_{\gamma\gamma} n'_t l'$, where
$\sigma_{\gamma\gamma}=(1/16)\sigma_{\rm T}$ is the cross section
for $\gamma\gamma$ absorption  ($\sigma_{\rm T}$ is the Thompson
cross section) and $n'_t$ is the  comoving-frame number density of
the target photons  that interact with high-energy photons. For
high energy photons with energy $E_h=5{\rm GeV}$, the energy of
the target photons is $\varepsilon_t\ga2(\Gamma m_e
c^2)^2/E_h=10{\rm keV} \Gamma_1^2 (E_h/5{\rm GeV})^{-1}$.   For a
power-law spectrum with a photon index of $\beta=2$ \citep{
2011Natur.476..421B}, the number density of target photons is
$n'_t\simeq L_X/(4\pi l'^2 \varepsilon'_t c \Gamma^4)$, where
$\varepsilon'_t=\varepsilon_t/\Gamma$. One can derive an upper
limit on the bulk Lorentz factor from the condition
$\tau_{\gamma\gamma}\ge 1$, i.e.
\begin{equation}
\Gamma\la \left(\frac{\sigma_{\gamma\gamma}L_X}{4\pi c^2 t_v
\varepsilon_t}
\right)^{1/4}=30L_{X,48}^{1/4}t_{v,2}^{-1/4}\left(\frac{\varepsilon_t}{\rm
10 keV}\right)^{-1/4}.
\end{equation}
This shows that the bulk Lorentz factors in TDEs are much lower
than that of GRBs \citep{2001ApJ...555..540L,2015ApJ...806..194T},
and may even be lower than  that of some blazars
\citep{2010A&A...512A..24S,2016MNRAS.455L...1W}. {There are other
possibilities  leading to non-detection of high energy emission
from these relativistic TDEs, such as cut-off in the spectrum of
accelerated electrons or a cooling break in the SED. More
multi-wavelength observations during flaring states will be
helpful to diagnose these different assumptions.}

\section{Conclusions}
We searched for the high-energy gamma-ray emission from TDEs with
the \textsl{Fermi} LAT survey data, including three  TDEs
confirmed with relativistic jets (Sw J1644+57, Sw J2058+05, Sw
J1112-82) and a nearby normal TDE ASASSN-14li. No significant
emission is found from these TDEs during the period from the
\textsl{Swift}/BAT trigger time (or the inferred disruption time)
to about one hundred days later. Compared with the bright
non-thermal X-ray emission,  the non-detection of high-energy
emission in three relativistic TDEs implies that high energy
emission is seriously suppressed, {possibly due to  the $\gamma
\gamma$ attenuation by soft photons in the source. Then, we derive
upper limits on the bulk Lorentz factors by assuming $\tau_{\gamma
\gamma} > 1$ for relativistic jets in TDEs. } The non-detection of
high-energy emission from the normal TDE ASASSN-14li gives an
upper limit  of $L(0.1-10{\rm GeV})\le 4.4\times 10^{42}{\rm erg}$
s$^{-1}$ during the first $10^{7}$ s after the disruption, which
is already lower than the X-ray luminosity
\citep{2016MNRAS.455.2918H}, but still higher than the predicted
gamma-ray luminosity \citep{2007A&A...473..351C}.

\section*{Acknowledgments}
This work made use of data supplied by the UK Swift Science Data
Center at the University of Leicester, and  Swift/BAT transient
monitor results provided by the Swift/BAT team. We thank Jin Zhang
for helpful discussion. This work is supported by the 973 program
under grant 2014CB845800, the NSFC under grants 11273016 and
11033002, and the Excellent Youth Foundation of Jiangsu Province
(BK2012011).

\clearpage

\begin{table}
\centering \caption{\textsl{Fermi} LAT  upper limit flux ($95\%$
confidence level) of four tidal disruption events at three time
intervals after their trigger time. Flux is in unit of erg
cm$^{-2}$ s$^{-1}$ (0.1-10 GeV).}
\begin{tabular}{lcccccccc}
\hline
\hline
Name&$\Gamma$& Time&Flux &Time&Flux & Time&Flux \\
&&s&$10^{-11}$ &s &$10^{-12}$ &s&$10^{-12}$ \\
\hline
Sw J1644+57&2.0 &0-$10^5$&2.41 &$10^5$-$10^6$& 7.63&$10^6$-$10^7$&1.18 \\
Sw J2058+05&2.0 &0-$10^5$&9.74 &$10^5$-$10^6$& 7.01&$10^6$-$10^7$&7.55 \\
Sw J1112-82&2.0 &0-$10^5$&5.51 &$10^5$-$10^6$& 25.7&$10^6$-$10^7$&5.75 \\
ASASSN\_14li & 2.0 &0-$10^5$&3.52 &$10^5$-$10^6$& 19.4&$10^6$-$10^7$&3.96 \\
\hline
\end{tabular}
\label{lc3bins}
\end{table}

\begin{table}
\centering \caption{Flux/Fluence of three TDEs and two blazars during their
flare states.  The unit of  flux is $10^{-8}$ ph cm$^{-2}$
s$^{-1}$, and the unit of the fluence is  $10^{-6}$ erg
cm$^{-2}$.}
\begin{tabular}{lcccccc}
\hline
\hline
& MJD & $\Gamma$& Fluence & Fluence & Fluence\\

&day& 0.1-10 GeV & 0.1-10 GeV  & 15-50 keV &10-1000 keV\\
\hline
Sw J1644+57&55648-55650&2.0&$<6.3$& $157.7\pm 10.0 $&506\\
Sw J2058+05&55698-55714&2.0&$<16.7$& $511.6\pm 118.3$ &1642\\
Sw J1112-82&55728-55731&2.0&$<9.0$&$299.2\pm 65.7$ &961\\
\hline
Mrk 421&55144-55149&$1.57\pm 0.21$&$95.5\pm 19.9$ &$638.6\pm 59.3$ &2050\\
&55242-55245&$1.46\pm 0.13$ & $135.3\pm 24.3$ &$341.0\pm 14.7$ &1095\\
&55246-55272&$1.81\pm 0.08$ &$397.0\pm 36.1$ &$1761.8\pm 48.9$ &5655\\
&55475-55503&$1.79\pm 0.06$ &$561.6\pm 41.9$ &$914.1\pm 68.2$ &2934\\
&55811-55818&$1.75\pm 0.11$ &$188.2\pm 27.0$ &$498.0\pm 47.0$ &1599\\
\hline
3C 279&54920-54980&$2.37\pm 0.03$ &$631.1\pm 6.94$ &$236.4\pm 116.5$&759\\
&55030-55050&$2.30\pm 0.01$ &$886.7\pm 9.66$ &$84.9\pm 68.0$ &273\\
\hline
\end{tabular}
\label{fluencecom}
\end{table}

\end{document}